\documentclass[useAMS,usenatbib]{mn2e}
\usepackage{graphicx}
\usepackage{epstopdf}

\def\Mpc{\hbox{$\rm\thinspace Mpc$}}
\def\kpc{\hbox{$\rm\thinspace kpc$}}

\def\hkpc{\hbox{$\thinspace h^{-1}\kpc$}}

\def\msun{\hbox{${\rm\thinspace M}_{\odot}$}}
\def\hrmsun{\hbox{$\thinspace h_{70}^{-1}\msun$}}

\title[The Limits of Bound Structures]{The Limits of Bound Structures in the Accelerating Universe}
\author[R. D\"unner et al.]{Rolando D\"unner$^1$, Pablo A. Araya$^2$, Andr\'es Meza$^3$\thanks{Researcher of the Academia Chilena de Ciencias 2004-2006} and Andreas Reisenegger$^1$ \\
$^1$Departamento de Astronom{\'\i}a y Astrof{\'\i}sica, Facultad de
F{\'\i}sica, Pontificia Universidad Cat\'olica de Chile,
Casilla 306, Santiago 22, Chile\\
$^2$Kapteyn Astronomical Institute, University of Groningen, P.O.
Box
800, 9700 AV Groningen, The Netherlands\\
$^3$Departamento de F\'{\i}sica, Facultad de Ciencias F\'{\i}sicas y Matem\'aticas, Universidad de Chile, Casilla 487-3, Santiago, Chile}

\begin{document}

\date{August 2005}

\pagerange{\pageref{firstpage}--\pageref{lastpage}} \pubyear{2005}

\maketitle

\label{firstpage}

\begin{abstract}
According to the latest evidence, the Universe is entering an era of
exponential expansion, where gravitationally bound structures will
get disconnected from each other, forming isolated `island
universes'. In this scenario, we present a theoretical criterion to
determine the boundaries of gravitationally bound structures and a
physically motivated definition of superclusters as the largest
bound structures in the Universe. We use the spherical collapse
model self-consistently to obtain an analytical condition for the
mean density enclosed by the last bound shell of the structure (2.36
times the critical density in the present Universe, assumed to be
flat, with 30 per cent matter and 70 per cent cosmological constant, in
agreement with the previous, numerical result of Chiueh and He).
$N$-body simulations extended to the future show that this
criterion, applied at the present cosmological epoch, defines a
sphere that encloses $\approx 99.7$ per cent of the particles that will
remain bound to the structure at least until the scale parameter of
the Universe is 100 times its present value. On the other hand,
$(28\pm 13)$ per cent of the enclosed particles are in fact not bound, so
the enclosed mass overestimates the bound mass, in contrast with the
previous, less rigorous criterion of, e.~g., Busha and
collaborators, which gave a more precise mass estimate. We also
verify that the spherical collapse model estimate for the radial
infall velocity of a shell enclosing a given mean density gives an
accurate prediction for the velocity profile of infalling particles,
down to very near the centre of the virialized core.
\end{abstract}


\section{Introduction}\label{intro}
The evidence for an accelerated expansion of the Universe, initially
based on the observations of distant supernovae \citep{SNCosmo,HiZ}
and later strengthened by precise measurements of cosmic microwave
background fluctuations \citep{WMAP}, has established a new
cosmological paradigm based on the presence of a `dark energy'
component. In the new scenario, the Universe has recently made a
(smooth) transition from a matter-dominated, decelerating stage to
a dark-energy dominated, accelerating stage.

In the simplest models, consistent with the observations so far, the
dark energy behaves like Einstein's cosmological constant, providing
an always present, constant, positive energy density (and a negative
pressure of the same magnitude). As long as the matter density was
substantially larger than the dark energy density, it dominated the
evolution of the Universe, decelerating the expansion and driving
the formation of structures by gravitational instability. When the
average matter density fell below that of the dark energy, the
latter started accelerating the expansion, and the formation of
structure slowed down, as the gravitational forces between matter
elements decreased due to their increasing separation. In this
stage, structures much denser than the dark energy are not affected
by the latter and remain bound, while they separate from each other
at an accelerating rate, which does not allow them to join in larger
structures. Thus, at the present cosmological time, when the
acceleration of the expansion has recently started, the largest
bound structures are just forming. In the future evolution of the
Universe, their individual, internal properties (such as physical
size and density) will not change substantially, but they will grow
increasingly isolated, forming `island universes' (e.~g.,
\citealt{Adams,Tzihong,Nagamine,Busha}).

In the present Universe, these structures have not yet fully formed,
virialized, and clearly separated from each other, making it
difficult to identify them unambiguously. Superclusters, the largest
structures identifiable in the present Universe, have generally been
defined by more or less arbitrary criteria (e.~g.,
\citealt{Quintana,Einasto0,Einasto1,Einasto2,Proust}). Here, we
address the question of how to decide whether such a structure will
remain gravitationally bound in the future evolution of the
Universe, forming an island universe, and propose to use this
criterion as a physical definition of superclusters.

We study the well-known spherical collapse model in the presence of
a cosmological constant, with the aim of obtaining a useful method
to study structure behavior in our Universe. The spherical collapse
model considers a spherically symmetric mass distribution, where
spherical shells expand or collapse into the centre of the structure
in a purely radial motion and without crossing each other.
Specifically, we study the density that needs to be enclosed by a
spherical shell at a given cosmological epoch in order to stay
gravitationally attached to the overdense region within in a
distant future, dominated by a cosmological constant.

Previously, \citet{Lokas2} gave an approximate criterion for this
density, assuming that the shell in question initially expands with
the Hubble flow (without retardation from the enclosed over-density)
and present density evolution curves considering different
cosmologies. \citet{Tzihong}, on the other hand, solved the
spherical collapse equations numerically, both with a cosmological
constant and with a more general form of dark energy (with a
constant `equation-of-state parameter' $w_q$), obtaining a
self-consistent, theoretical criterion for the mean density enclosed
in the last gravitationally bound or `critical' shell. Both
\citet{Nagamine} and \citet{Busha} studied the future evolution of
structures numerically, contrasting the extension of bound
structures in a distant future (at scale parameter $a=166$ and
$a=100$, respectively) with the criterion of \cite{Lokas2} applied
at the present time ($a=1$). While \citet{Nagamine} focused on the
evolution of specific structures in our local Universe (the Local
Group, the Virgo cluster, and other nearby structures),
\citet{Busha} followed the internal density and velocity structures
of generic, bound objects as they evolve, a subject taken up again
by the same authors more recently \citep{Busha05}.

We stress that, among the previous work cited above, only
\citet{Tzihong} used the spherical collapse model self-consistently
in order to obtain the density criterion for the last bound shell,
whereas all the other papers rely on the incorrect assumption that
the shell expands with the Hubble flow at the initial epoch (taken
to be the present time, both by \citealt{Nagamine} and
\citealt{Busha}). On the other hand, \citet{Tzihong} did not perform
numerical simulations of the future evolution of structure in order
to test the accuracy of their result in a non-ideal situation and in
order to study the details of the evolution of bound objects, as the
other authors did. In this work, we attempt to bring together and
extend the best parts of the previous work, by a self-consistent
application of the spherical collapse model (supplemented by a new,
analytic equation for the mean density of a marginally bound sphere)
and a comparison of its predictions to $N$-body simulations of the
future Universe.

In section \ref{Spherical Collapse Model}, we review the spherical
collapse model, deriving an analytical solution for the spherical
collapse equations. Our solution, which agrees with the numerical
result of \citet{Tzihong}, relates the critical shell's overdensity
with the value of the dimensionless parameter
$\Omega_{\Lambda}(t)=\Lambda/[3H(t)^2]$ (where $\Lambda$ is the
cosmological constant and $H(t)$ is the Hubble parameter), so we are
able to evaluate the criterion at any time $t$ of the evolution of
the Universe.
We also obtain numerical solutions for the velocities of
non-critical shells in the present Universe
($\Omega_{\Lambda}=0.7$).

In sections \ref{Numerical Simulations} through \ref{Comparison
Between Theory and Simulated Data}, we compare the theoretical
results with simulated data, studying the quality of the criterion
and its applicability to real-world observations. As in
\citet{Nagamine} and \citet{Busha}, \mbox{$N$-body} simulations were
run until the very late future ($a=100$), in order to reproduce the
final configuration of bound structures. We find that the rigorous
density criterion proposed by us (in agreement with
\citealt{Tzihong}) gives a good upper bound to the size of the bound
structure, as it encloses $\approx 99.7$ per cent of the bound particles.
On the other hand, it also encloses a substantial number of unbound
particles, and therefore does not perform as well as the criterion
of \citet{Lokas2} (used by \citealt{Nagamine} and \citealt{Busha})
for the purpose of estimating the bound mass. We also show that the
spherical collapse model is quite accurate in predicting (as a
function of the enclosed mean density) the radial velocity of the
stream of particles falling into a bound structure for the first
time, down to the very centre of the structure.

\section{Spherical collapse model}
\label{Spherical Collapse Model} A straightforward approach to the
evolution of structure is obtained by considering a spherical
distribution where all layers expand or contract with only radial
motions and without crossing each other. The later ensures that the
enclosed mass in every shell stays constant throughout its whole
evolution, being the only parameter other than initial conditions to
determine its behavior.
The model is more accurate during the expansion phase, but becomes unrealistic after
contraction begins, because the instability in angular momentum
during this phase causes non-radial motions and finally
virialization. In our analysis, we are going to leave aside this
warning and check the accuracy of the model in the contraction phase
using comparison with simulations.

Our analysis will be based on Newtonian Mechanics, which, according
to \citet{Lemaitre}, is a limiting approximation to general
relativity, valid no matter what is happening in distant parts of
the Universe. The Newtonian approximation is accurate in a region
small compared to the Hubble length $c/H$ and large compared to the
Schwarzschild radii of any collapsed objects. For more details, see
\citet{Peebles1980}.

Under the assumption that the total energy will be conserved during the shell's expansion and later contraction, the evolution of a spherical shell enclosing a spherically symmetric mass distribution is given by the energy equation \citep{Peebles1980}:

\begin{equation}
\label{ } E={1\over 2}\left(dr\over
dt\right)^2-\frac{GM}{r}-\frac{\Lambda}{6}r^{2},
\end{equation}
where $r$ is the shell's radius, $M$ is the total mass enclosed by
it, $\Lambda$ is the cosmological constant and $E$ is the total
energy per unit mass of the shell.

This equation can be simplified by introducing the following dimensionless variables:
\begin{equation}
\label{eq:r_barra}
\widetilde{r}=\left(\frac{\Lambda}{3GM}\right)^{1/3}r,\\
\end{equation}
\begin{equation}
\label{ }
\widetilde{t}=\left(\frac{\Lambda}{3}\right)^{1/2}t .\\
\end{equation}
Therefore, the equation may be written
\begin{equation}
\widetilde{E}={1\over 2}\left(d\widetilde{r}\over d\widetilde
t\right)^{2}-\frac{1}{\widetilde{r}}-\frac{\widetilde{r}^{2}}{2},
\label{eq:E_b}
\end{equation}
where
\begin{equation}
\label{ }
\widetilde{E}=E \left(\frac{G^{2}M^{2}\Lambda}{3}\right)^{-1/3}.
\end{equation}

Fixing $M$ and considering that there is no shell crossing, this equation describes the time evolution of a single shell. $\widetilde{E}$ is the dimensionless energy that describes every shell of the distribution, merging with the background's mean density when $\widetilde{E}=0$, shell that will expand with the Hubble flow.

\subsection{The critical shell and turn-around radius}

We are looking for a critical shell that will stay at the limit between expanding for ever or re-collapsing into the structure. To find the critical energy for such a shell, we define a potential energy to be maximized as
\begin{equation}
\label{ }
\widetilde{V} =- \frac{1}{\widetilde{r}}-\frac{\widetilde{r}^{2}}{2}.
\end{equation}
The maximum of this potential occurs at $\widetilde{r}^* = 1$, so
$\widetilde{E}^*\equiv\widetilde{V}\left(\widetilde{r}^*\right)=-\frac{3}{2}$
is the maximum possible energy for a shell to remain attached to the
structure. The maximum radius for a critical shell is
\begin{equation}
\label{ }
r_\mathrm{max}\equiv\left(\frac{3GM}{\Lambda}\right)^{1/3},
\end{equation}
so we can reinterpret the normalized radius $\widetilde{r}$ as
\begin{equation}
\label{ } \widetilde{r}\equiv\frac{r}{r_\mathrm{max}}.
\end{equation}



\subsection{Connection with the background model}

Assuming a flat universe with cosmological constant, the age of the
Universe (time since the Big Bang, written in our dimensionless
variables) may be related to the vacuum energy density parameter
\citep{Peebles1980},
\begin{equation}
\label{eq:t_Omega} \Omega_\Lambda\equiv{\Lambda\over 3
H^2}=\tanh^2(3\widetilde{t}/2),
\end{equation}
showing that $\Omega_{\Lambda}$ increases monotonically in time, and
can therefore be used as a surrogate time variable, in terms of
which we will describe the evolution of bound or critical
(marginally bound) spherical shells.

Integrating equation (\ref{eq:E_b}) from the beginning of time
($\widetilde{r} = 0$) till the current radius of a given shell
($\widetilde{r} = \widetilde{r}_{0}$), we get
\begin{equation}
\label{eq:t_int}
\widetilde{t}_{0}=\int^{\widetilde{r}_{0}}_{0}\frac{\sqrt{\widetilde{r}}d\widetilde{r}}{\sqrt{\widetilde{r}^{3}+2\widetilde{E}\widetilde{r}+2}}.
\end{equation}
In the particular case of a critical shell
($\widetilde{E}=\widetilde{E}^*=-\frac{3}{2}$), the denominator of
equation (\ref{eq:t_int}) can be factorized, yielding
\begin{equation}
\label{eq:t_int_c} \widetilde{t}_{0}=\int
^{\widetilde{r}_{0}}_{0}\frac{\sqrt{\widetilde{r}}d\widetilde{r}}{\left(1-\widetilde{r}\right)\sqrt{\widetilde{r}+2}}.
\end{equation}
The integral can be done analytically, with the result
\begin{eqnarray}
\label{eq:sol_int}
\widetilde{t_{0}}=\frac{1}{2\sqrt{3}}\ln\left[\frac{1+2\widetilde{r_{0}}+\sqrt{3\widetilde{r_{0}}\left(\widetilde{r_{0}}+2\right)}}{1+2\widetilde{r_{0}}-\sqrt{3\widetilde{r_{0}}
\left(\widetilde{r_{0}}+2\right)}}\right]\\
-\ln\left[1+\widetilde{r_{0}}+\sqrt{\widetilde{r_{0}}\left(\widetilde{r_{0}}+2\right)}\right]. \nonumber
\end{eqnarray}

Notice that $\widetilde{r}_0$ can be chosen at any `current' time,
so equation (\ref{eq:sol_int}) can be generalized to any time by
simply dropping subscript 0. From now on, we will use the subscript
`cs' to indicate that we are referring to a critical shell.

\begin{figure}
    \flushright
    \includegraphics[trim=0 -15 0 0 , width=80mm]{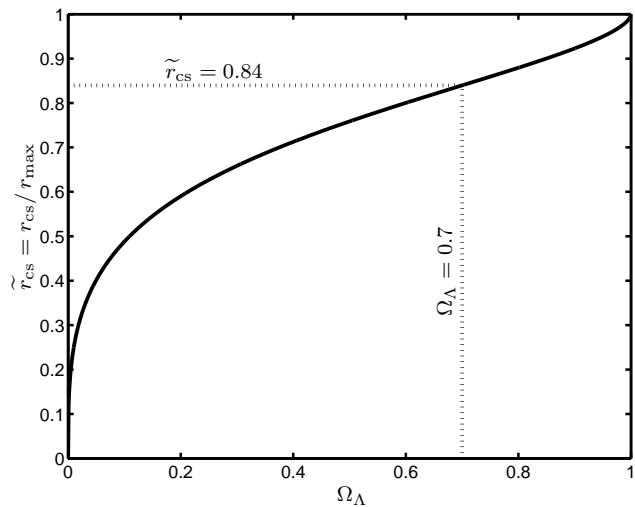}
    \put(-115,0){$\Omega_{\Lambda}$}
    \put(-238,80){\rotatebox{90}{$\widetilde{r}_\mathrm{cs}=r_\mathrm{cs}/\,r_\mathrm{max}$}}
    \put(-77,70){\rotatebox{90}{$\Omega_{\Lambda}=0.7$}}
    \put(-180,160){$\widetilde{r}_\mathrm{cs}=0.84$}
    \caption{The normalized radius $\widetilde{r}_\mathrm{cs}$ of a
    critical shell as a function of $\Omega_{\Lambda}$. The dotted
    lines highlight the present universe, in which $\Omega_{\Lambda}=0.7$
    and $\widetilde{r}_\mathrm{cs}=0.84$.}
    \label{fig:Om_r}
\end{figure}

Replacing eq. (\ref{eq:sol_int}) in eq.(\ref{eq:t_Omega}), and
introducing a new variable
\begin{eqnarray} \label{ }
\chi\left(\widetilde{r}_\mathrm{cs}\right) \equiv
\left[\frac{1+2\widetilde{r}_\mathrm{cs}+\sqrt{3\widetilde{r}_\mathrm{cs}\left(
\widetilde{r}_\mathrm{cs}+2\right)}}{1+2\widetilde{r}_\mathrm{cs}-\sqrt{3\widetilde{r}_\mathrm{cs}
\left(\widetilde{r}_\mathrm{cs}+2\right)}}\right]^{\frac{\sqrt{3}}{2}}\\
\times\left(1+\widetilde{r}_\mathrm{cs}+\sqrt{\widetilde{r}_\mathrm{cs}\left(\widetilde{r}_\mathrm{cs}+2\right)}\right)^{-3},\nonumber
\end{eqnarray}
we can write the relation between $\Omega_{\Lambda}$ (and therefore
cosmological time) and $\widetilde{r}_\mathrm{cs}$ as
\begin{equation}
\label{eq:Omega_r}
\Omega_{\Lambda}\left(\widetilde{r}_\mathrm{cs}\right)=\left[\frac{\chi\left(\widetilde{r}_\mathrm{cs}\right)-1}{\chi\left(\widetilde{r}_\mathrm{cs}\right)+1}\right]^{2}.
\end{equation}
As expected (see Fig. \ref{fig:Om_r}), $\widetilde{r}_\mathrm{cs}$
grows with $\Omega_{\Lambda}$ as $\Omega_{\Lambda}$ grows with time,
and will converge to its maximum radius when $t\rightarrow \infty$.

Inversely, evaluating equation (\ref{eq:Omega_r}) at our preferred
cosmology ($\Omega_\mathrm{m} = 0.3$, $\Omega_\Lambda = 0.7$), we
obtain that current critical shells should have
$\widetilde{r}_\mathrm{cs}=0.84$, which means that their present
radius is 84 per cent of the maximum radius they will reach as
$t\rightarrow \infty$.

\subsection{Conditions for a critical shell}

For practical applications, it is convenient to express the critical
condition as the minimum enclosed mean density needed by a shell to
stay bound to the central gravitational attractor. The critical
density of the Universe (in the flat model, also the average total
density, including matter and vacuum energy) is
\begin{equation}
\label{ } \rho_\mathrm{c} = \frac{3H^{2}}{8\pi G},
\end{equation}
and the average mass density enclosed by a given shell is
\begin{equation}
\bar{\rho}_\mathrm{m}^\mathrm{s}=\frac{3M}{4\pi r^3}.
\end{equation}
Defining the mass density parameter for the shell,
\begin{equation}
\label{eq:F}
\Omega_\mathrm{s}\equiv\frac{\bar{\rho}_\mathrm{m}^\mathrm{s}}{\rho_\mathrm{c}}=\frac{2\Omega_{\Lambda}}{\widetilde{r}^3},
\end{equation}
the condition for the shell to be bound is
%
\begin{equation}
\label{eq:Fc}
\Omega_\mathrm{s}\geq\Omega_\mathrm{cs}=\frac{2\Omega_{\Lambda}}{\widetilde{r}_\mathrm{cs}^{3}}
= 2.36,
\end{equation}
where we have evaluated equation (\ref{eq:Omega_r}) at the present
value of $\Omega_\Lambda = 0.7$. This represents the present density
contrast inside the last shell that will eventually stop its growth
at the end of times. This result was first obtained by
\citet{Tzihong}, using numerical methods, but now we confirm this
result with an analytical solution. In the same way, evaluating at
$\Omega_\Lambda = 1$ ($t\rightarrow \infty$) we obtain the
asymptotic critical density contrast condition
$\Omega_{\mathrm{cs},\infty} = 2$.

\begin{figure}
    \flushright
    \includegraphics[trim=0 -15 0 -20 , width = 80mm]{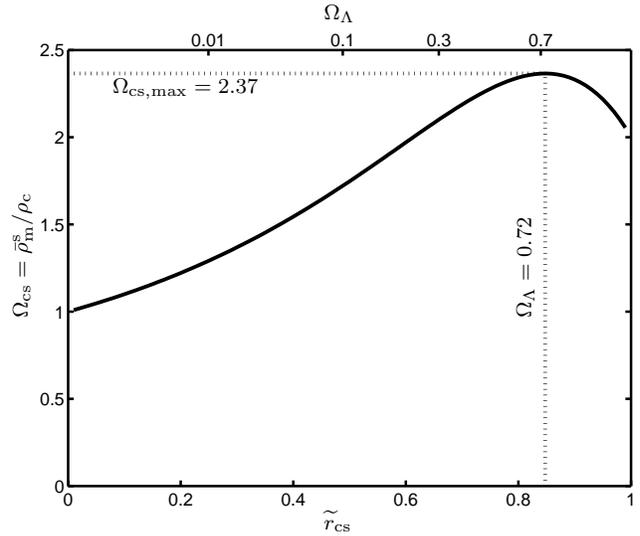}
    \put(-120,0){$\widetilde{r}_\mathrm{cs}$}
    \put(-120,193){$\Omega_{\Lambda}$}
    \put(-238,80){\rotatebox{90}{$\Omega_\mathrm{cs}=\bar{\rho}_\mathrm{m}^\mathrm{s}/\rho_\mathrm{c}$}}
    \put(-47,80){\rotatebox{90}{$\Omega_{\Lambda}=0.72$}}
    \put(-200,165){$\Omega_\mathrm{cs,max}=2.37$}
    \caption{Density parameter for a critical shell, $\Omega_\mathrm{s}$ (see equation (\ref{eq:F})), as a
    function of $\Omega_{\Lambda}$ and dimensionless radius $\widetilde{r}_\mathrm{cs}$. Its maximum value,
    $\Omega_\mathrm{cs}=2.37$,
    occurs at $\Omega_{\Lambda}=0.72$ ($\widetilde{r}_\mathrm{cs}=0.85$).}
    \label{fig:F}
\end{figure}

Fig. \ref{fig:F} shows the density parameter $\Omega_\mathrm{cs}$ as
a function of $\Omega_{\Lambda}$ and $\widetilde{r}_\mathrm{cs}$
using equation (\ref{eq:Omega_r}) or its inverse. It is curious that
the curve shows a single maximum at $\Omega_{\Lambda}=0.72$, very
close to the measured value of $\Omega_{\Lambda}$ today. Considering
that $\Omega_{\Lambda}$ can be taken as a function of time, this
means that we are living in the era when the ratio between mass
density inside a critical shell and the critical density of the
Universe is very nearly at its maximum.

For observational purposes, it is more interesting to know the ratio
between the mass density enclosed by the critical shell,
$\bar{\rho}^\mathrm{s}_\mathrm{m}$, and that of the background,
$\rho^\mathrm{b}_\mathrm{m}$. Using equation (\ref{eq:F}), we obtain
that
\begin{equation}
\label{eq:ratio_m_m}
\frac{\bar{\rho}^\mathrm{s}_\mathrm{m}}{\rho^\mathrm{b}_\mathrm{m}}
=\frac{\bar{\rho}^\mathrm{s}_\mathrm{m}}{\rho_\mathrm{c}\left(1-\Omega_{\Lambda}\right)}
=\frac{\Omega_\mathrm{cs}}{1-\Omega_{\Lambda}}.
\end{equation}
which, evaluated for $\Omega_{\Lambda}=0.7$, yields
\begin{displaymath}
\frac{\bar{\rho}^\mathrm{s}_\mathrm{m}}{\rho^\mathrm{b}_\mathrm{m}}=7.88.
\end{displaymath}

In the notation of \citet{Busha}, considering only the excess
density with respect to the background ($M_\mathrm{obj} = M -
M_\mathrm{b}$, where $M_\mathrm{b}$ is the amount of mass contained
in an equivalent sphere with background density) and writing the
present value of the Hubble parameter as
$H_0=70~h_{70}\mathrm{km~s^{-1}~Mpc^{-1}}$, we rewrite the criterion
as
\begin{equation}
\label{ }
\frac{M_\mathrm{obj}}{10^{12}M_{\odot}}\geq1.18~h_{70}^2\left(\frac{r_{0}}{1\Mpc}\right)^3.
\end{equation}
This is a less restrictive condition than the one proposed by
\citet{Busha}, which is
\begin{equation}
\label{ } \frac{M_\mathrm{obj}}{10^{12}M_{\odot}}\geq
3~h_{70}^2\left(\frac{r_{0}}{1\Mpc}\right)^3,
\end{equation}
or, equivalently, $\Omega_\mathrm{s} \geq 5.56$ (also equivalent to
the overdensity criterion given by \citealt{Nagamine}, based on the
formalism of \citealt{Lokas2}). This result was expected, since the
criterion of \citet{Busha} was obtained under the assumption that
the test particles, placed on the critical shell today, move with
the Hubble flow. As this is not really true, they free that
parameter (velocity of particles on the critical shell today) and do
an empirical test to observe the real behavior of particles evolved
till $a = 100$. They obtain a corrected result which is closer to
their initial result than to ours, always setting a higher
constraint to the critical shell density. A probable reason for
this difference is that our result is a completely theoretical
approach based on the spherical collapse model (only radial
motions), but in real life objects usually obtain angular momentum
caused by tidal forces, which will tend to detach objects from the
structure, strengthening the binding condition. This discussion will
be resumed later, in relation to the results of our simulations.

\subsection{Velocities of shells}
\label{Velocity of Non-critical Shells} In order to work with shell
velocities, it is convenient to refer the radial velocities to the
Hubble flow, so we introduce the parameter $A$, defined as in
\citet{Busha},
\begin{equation}
\label{eq:A} A\left(\Omega_{\Lambda}\right)\equiv\left({1\over
H_0r}{dr\over dt}\right)^{2}.
\end{equation}
For a critical shell, we can use equation (\ref{eq:E_b}) to write this in terms of $\Omega_{\Lambda}$ as
\begin{equation}
\label{eq:A_cs}
A_\mathrm{cs}\left(\Omega_{\Lambda}\right)=\Omega_{\Lambda}\left[
1-\frac{3}{\widetilde{r}^{2}\left(\Omega_{\Lambda}\right)}+\frac{2}{\widetilde{r}^{3}\left(\Omega_{\Lambda}\right)}\right].
\end{equation}
Evaluating at $\Omega_\Lambda = 0.7$, we obtain that the present
velocity parameter for a critical shell is
$A_\mathrm{cs}=8.63\times10^{-2}$, showing that the shell has been
slowed down substantially with respect to the Hubble flow.

\begin{figure}
    \flushright
    \includegraphics[trim=0 -15 0 0 , width = 80mm]{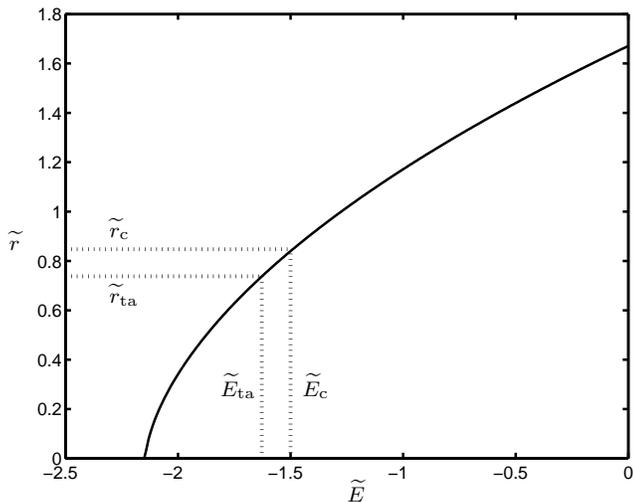}
    \put(-110,0){$\widetilde{E}$}
    \put(-238,95){$\widetilde{r}$}
    \put(-158,40){$\widetilde{E}_\mathrm{ta}$}
    \put(-127,40){$\widetilde{E}_\mathrm{c}$}
    \put(-200,75){$\widetilde{r}_\mathrm{ta}$}
    \put(-200,100){$\widetilde{r}_\mathrm{c}$}
    \caption{The dimensionless radius $\widetilde{r}$ of a shell as a function of its dimensionless energy
    $\widetilde{E}$ at a cosmological time characterized by $\Omega_{\Lambda} = 0.7$.
    A shell with $\widetilde{E}=-2.1$ corresponds to $\widetilde{r}=0$, so it is now collapsing.
    Highlighted by dotted vertical and horizontal lines are the critical shell, with $\widetilde{E}_\mathrm{c}=-1.5$ and $\widetilde{r}_\mathrm{c}=0.84$, and the shell at turn-around, with $\widetilde{E}_\mathrm{ta}=-1.64$ and $\widetilde{r}_\mathrm{ta}=0.73$.}
    \label{fig:r_E}
\end{figure}

To study the velocity profile of a mass overdensity, it is
convenient to relate the energy of an arbitrary shell to its
normalized radius, characterized by the present value of
$\Omega_{\Lambda}$. Unfortunately, the energy integral has no
analytical solution for $\widetilde{E} \neq \widetilde{E}^*$, so
this relation can only be obtained numerically. Fixing the value of
$\Omega_{\Lambda}$ to 0.7, in order to represent the present
universe, we obtain the current normalized time
$\widetilde{t}_{0}=0.81$. Then, we numerically integrate equation
(\ref{eq:t_int}) for every possible value of $\widetilde{E}$ from
zero till some value of $\widetilde{r}$ that satisfies the time
constraint. Now using equation (\ref{eq:F}), we arrive at the more
useful numerical function $\widetilde{E}(\Omega_\mathrm{s};
\Omega_\Lambda)$, which is best shown in Fig. \ref{fig:r_E}.

Expressing the velocity parameter $A$ in terms of an arbitrary
normalized energy $\widetilde{E}(\Omega_\mathrm{s};
\Omega_\Lambda)$, for a fixed value $\Omega_\Lambda$, we obtain

\begin{equation}
\label{eq:v_A} A =
\Omega_\Lambda+\Omega_\mathrm{s}+\widetilde{E}(\Omega_\mathrm{s};
\Omega_\Lambda)\left(2\Omega_\Lambda\Omega_\mathrm{s}^2\right)^{1/3}.
\end{equation}

\begin{figure}
    \flushright
    \includegraphics[trim=0 -15 0 0 , width = 80mm]{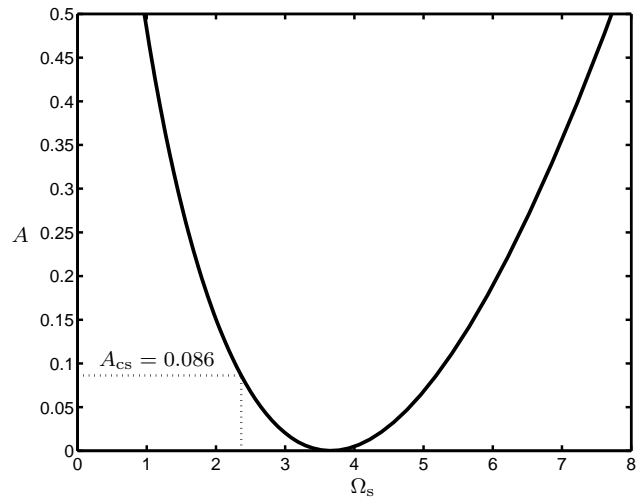}
    \put(-110,0){$\Omega_\mathrm{s}$}
    \put(-238,95){$A$}
    \put(-205,48){$A_\mathrm{cs}=0.086$}
    \caption{Radial velocity parameter $A$ (see eq. \ref{eq:v_A}) for a shell as a function of the
    enclosed density $\Omega_\mathrm{s}$ (eq. \ref{eq:F}) in a universe with $\Omega_\Lambda=0.7$.
    For a critical shell, $\Omega_\mathrm{cs}=2.36$ and $A(\Omega_\Lambda)=0.086$.
    For a shell at its turn-around point, $A=0$ and $\Omega_\mathrm{s}=3.66$.}
    \label{fig:A_OmegaM}
\end{figure}

In Fig. \ref{fig:A_OmegaM}, we observe that $A(\Omega_\mathrm{s})$
is zero when $\Omega_\mathrm{s} = \Omega_\mathrm{s,ta}$, the density
parameter corresponding to the turn-around radius. When solving for
the radial velocity, it is important to notice that shells with
$\Omega_\mathrm{s}>\Omega_\mathrm{s,ta}$ contract, while shells with
$\Omega_\mathrm{s}<\Omega_\mathrm{s,ta}$ expand. For a universe with
$\Omega_\Lambda=0.7$, $\Omega_\mathrm{s,ta}=3.66$.

\section{Numerical simulations}
\label{Numerical Simulations} We simulate one cosmological model,
assuming a standard flat Lambda Cold Dark Matter ($\Lambda$CDM)
Universe. The current cosmological parameters in the simulation are
$\Omega_{\mathrm{m},0} =0.3$, $\Omega_{\Lambda,0}=0.7$ and $h=0.7$,
where the Hubble parameter $H_{0}=100h$ km s$^{-1}$Mpc$^{-1}$. The
normalization of the power spectrum is $\sigma_{8}=1$. The box has a
side length of 100$h^{-1}$Mpc and contains 128$^{3}$ dark matter
particles of mass $m_\mathrm{DM}=3.97\times 10^{10}h^{-1}M_{\odot}$.
The simulation was evolved from $a=0.02$ (redshift $z=49$) to
$a=100$. The Plummer-equivalent gravitational softening was set to
$\epsilon_\mathrm{Pl}=15$ $\hkpc$ (physical units) from $a = 1/3$ to $a=100$, while it was taken to be fixed in comoving units at
higher redshift.

The run was performed with the massive parallel tree $N$-body/SPH
code \begin{footnotesize}GADGET\end{footnotesize} \citep{Springel}.
This is a \begin{footnotesize}TREESPH\end{footnotesize} code where
the dark matter particles are evolved using a tree-code, while the
collisional gas is followed using the SPH approach\footnote{The gas
particles were not considered in our simulations.}. Here, we used the
new improved
\begin{footnotesize}GADGET2\end{footnotesize}, kindly provided by
Volker Springel \citep{Springel2}, which is more memory-efficient,
offers better time-stepping for collisionless dynamics, and is
substantially faster than the original version. The initial
conditions were established by the code of \citet{Weygaert}.

We took snapshots at the present time ($a=1$) and in the far future
($a=100$), assuming that in late epochs the structure evolution will
decrease significantly, so no major changes will be seen from then
on. The identification of structures was done using a
friend-of-friends code to identify the initial candidates and later
extracting the structures to produce a reduced catalog for our
study.

\section{Application of binding criteria to simulation data}

We identified 22 mass concentrations with masses greater than $10^{14}~\hrmsun$,
from which we selected 11 that were away from the box boundaries in
order to do our analysis\footnote{We could have used the other 11
objects as well, because the box had periodic boundary conditions,
but we did not, just to simplify the analysis.}. The same
identification was done in the future frame ($a=100$), obtaining the
final state for the structures identified at $a = 1$. We were able
to follow particles from the present till the late future frame, so
we could exactly determine the fate of every particle.

\begin{figure}
    \flushright
    \includegraphics[trim=0 -15 0 0 , width = 80mm]{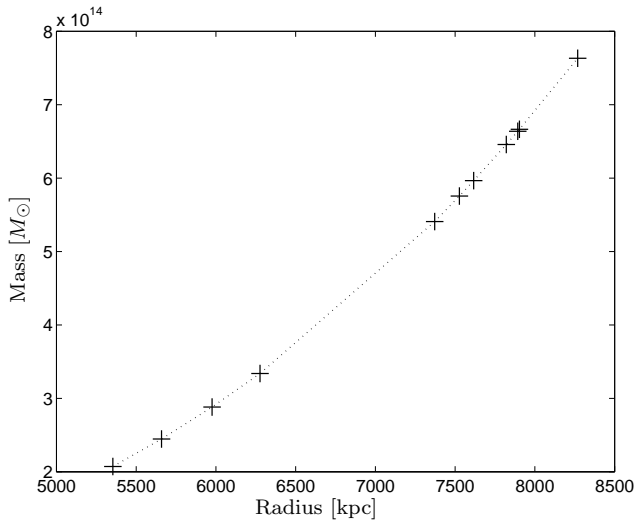}
    \put(-145,0){Radius [kpc]}
    \put(-238,80){\rotatebox{90}{Mass $[M_{\odot}]$}}
    \caption{Enclosed mass as a function of critical radius for 11
    bound objects identified in the simulation at $a=1$. The dotted
    line represents the constraint of 2.36 times the critical density,
    used to define the bound region.}
    \label{fig:MvsR}
\end{figure}

In order to apply the density criterion for gravitational binding,
we first identified the densest core of the structure. For this, we
chose a centre and found the radius where the mean inner density was
300 times the critical density of the Universe (chosen to identify a
dense, virialized core that would be clearly detectable in
observations). We then re-centred the sphere to its centre of mass.
We repeated this procedure until the centre of mass matched the
geometric centre of the sphere. Once the centre was fixed, we
calculated the density parameter $\Omega_\mathrm{s}$ of concentric
spheres with increasing radius until the condition
$\Omega_\mathrm{s} < \Omega_\mathrm{cs} = 2.36$ stopped being
satisfied. We called the identified radius $r_\mathrm{c}$. (The
radii and masses of these structures are plotted in Fig.
\ref{fig:MvsR}). The same identification was done at $a = 100$, with
the critical condition $\Omega_\mathrm{s} <
\Omega_{\mathrm{cs},\infty} = 2$. In order to contrast our results,
we applied the criterion of \citet{Busha} at $a = 1$, which is given
by $\Omega_\mathrm{s} < \Omega_\mathrm{cs,B} = 5.56$. We called the
identified radius $r_B$.

\begin{figure}
    \flushright
    \includegraphics[trim=0 -15 0 0 , width = 80mm]{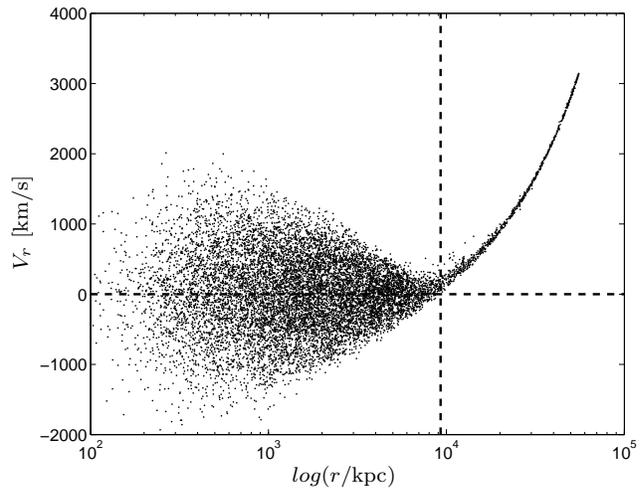}
    \put(-132,0){$log(r / \rm{kpc})$}
    \put(-238,80){\rotatebox{90}{$V_r$ [km/s]}}
    \caption{Radial velocities as a function of radius (in
    physical units) for a mass overdensity at $a=100$. The vertical dashed
    line shows the radius where  $\Omega_\mathrm{s} = \Omega_{\mathrm{cs},\infty}= 2$.}
    \label{fig:vr_100}
\end{figure}

A critical part of the analysis is to have an acceptable criterion
to determine whether a particle is bound to the overdense structure
or not. As illustrated in Fig. \ref{fig:vr_100}, the theoretical
criterion at $a=100$ gives a very intuitive result, placed at the
end of the virialized region, almost exactly where the lower
envelope of the radial velocities crosses zero. In other words,
$\Omega_{\mathrm{cs},\infty}$ coincides with the last radius where
objects with negative radial velocity can be found. At that point,
only a few particles will be able to escape, so we can say that the
criterion at $a=100$ is adequate to determine the limits of bound
structures in the distant future universe. A very similar figure, as
well as a detailed discussion of how the velocity distribution
evolves to this state, was recently given by \citet{Busha05}.

Finally, we classified particles in four categories: particles that
fall inside the theoretical critical shell at $a = 1$ and $a = 100$;
particles that fall inside the critical shell at $a = 1$ but do not
at $a = 100$; particles that fall outside the critical shell at $a =
1$ but inside at $a = 100$ and particles that fall outside in both
cases. This categorization lets us visualize the quality of our
estimation, clearly separating particles according to their real
fate, and permitting us to calculate statistical indicators useful
to produce the desired error estimations. As expected, not every
object presented a very `spherical' distribution at $a = 1$, and
this affected how well data were fitted by the spherical
model\footnote{We even found some objects that were currently
undergoing mergers, so they showed strong evolution between $a = 1$
and $a = 100$.}. For this reason, we based our qualitative analysis
on the `best' objects, but we kept all of them when doing
statistics.

\section{Comparison between theory and simulated data}
\label{Comparison Between Theory and Simulated Data} The spherical
model predicts a purely radial motion of particles towards the
centre. This is clearly not the case in the real world, where
objects are affected by multiple and complicated tidal forces
produced by other objects in their surroundings all along their
evolution. In fact, objects present a velocity dispersion that is
greatest in the virialized cores of galaxy clusters. The presence of
other overdensities surrounding the main attractor alters the
motion, pushing particles away from their radial trajectories. For
this reason, we expect the spherical model to give a lower bound on
the radial velocities at a given radius (with outward-pointing
velocities taken as positive) and an upper limit for the radius of
critical shells.
\begin{figure*}
    \includegraphics[width = 80mm]{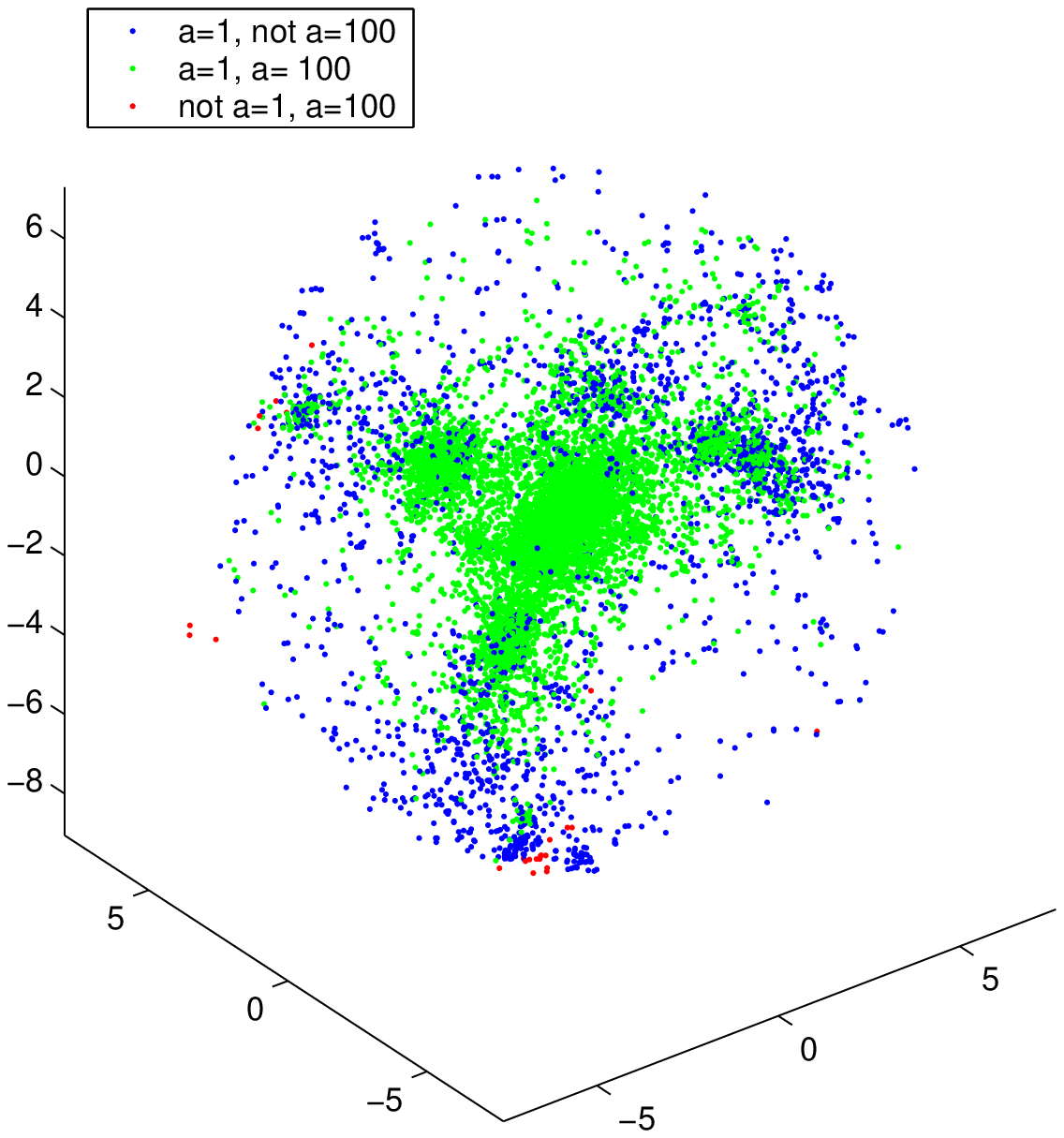}
    \includegraphics[width = 80mm]{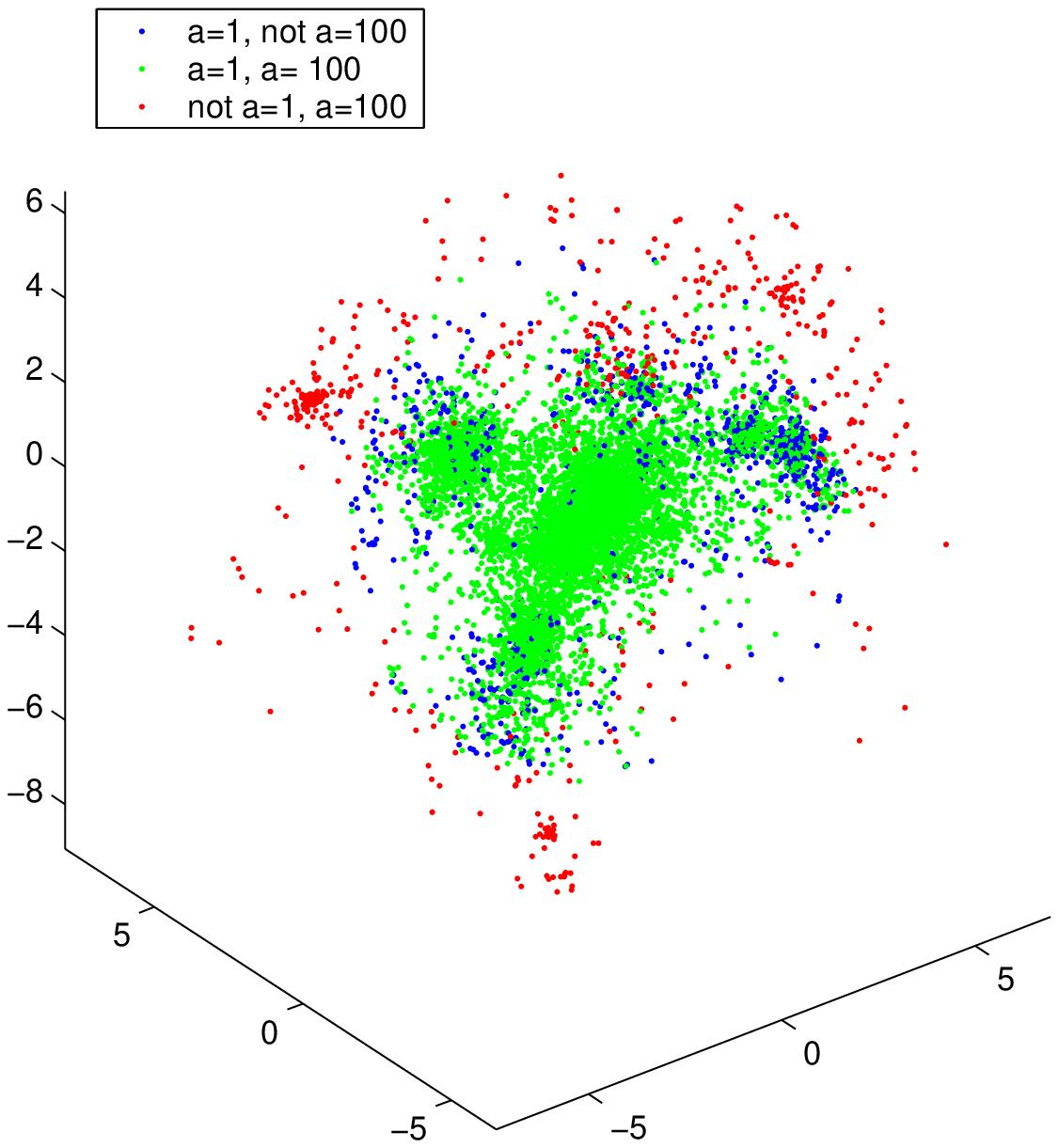}
    \put(-205,20){$x$ [Mpc]}
    \put(-437,20){$x$ [Mpc]}
    \put(-50,10){$y$ [Mpc]}
    \put(-280,10){$y$ [Mpc]}
    \put(-238,120){\rotatebox{90}{$z$ [Mpc]}}
    \put(-470,120){\rotatebox{90}{$z$ [Mpc]}}
    \caption{Spatial distribution of particles in object no. 1 (see Table \ref{tabla_D}) with colours depending on our criterion (left-hand panel), or on that of \citet{Busha} (right-hand panel). Blue: they satisfy the criterion at $a = 1$, but not at $a = 100$, Green: they satisfy the criterion at $a = 1$ and $a = 100$, Red: they do not satisfy the criterion at $a = 1$, but they do at $a = 100$. Particles that do not satisfy the criterion at either $a=1$ or $a=100$ are not shown.}
    \label{fig:3D_distr}
\end{figure*}

To test the performance of the criterion, we first selected all the
particles that were bound at $a=100$ and `marked' them, so they
could be recognized at $a=1$, as shown for the most massive
structure in Fig. \ref{fig:3D_distr}. Later, we selected all the
particles that satisfied the criterion at $a=1$ and produced four
statistical indicators\footnote{Only two of these are independent,
as $A+B=1$ (or 100 per cent) and $B+C=D$.}: $A$, the fraction of all
particles selected by the criterion as bound at $a=1$ that were not
selected at $a=100$; $B$, the fraction of all particles selected at
$a=1$ that were also selected at $a=100$; $C$, the ratio of the
number of particles not selected at $a=1$ but selected at $a=100$, to
the total number of selected particles at $a=1$ and $D$, the ratio
of the mass still bound at $a = 100$ to the mass selected at $a =
1$. Results were averaged over all objects to obtain a single
indicator. Results for our criterion are shown in Table
\ref{tabla_D} and for that of \citet{Busha} in Table \ref{tabla_B}.

\begin{table*}
  \caption{Quantitative results after application of our theoretical
  criterion to 11 objects from the simulation. Columns: (1) index of the object;
  (2) $A$, particles selected by the criterion as bound at $a=1$ but not at $a=100$,
  as fraction of all particles selected at $a=1$; (3) $B$, particles selected at $a=1$
  and $a=100$, as fraction of all particles selected at $a=1$ (note $A+B=100$ per cent);
  (4) $C$, ratio of the number of particles not selected at $a=1$ but selected at
  $a=100$, to the total number of selected particles at $a=1$;
  (5) $M(a=1)$, total mass selected inside the critical radius at $a=1$;
  (6) $M(a=100)$ total mass selected inside the critical radius at $a=100$ and
  (7) $D=M(a=100)/M(a=1).$}
  \begin{small}
  \begin{tabular}{| c | c | c | c | c | c | c |}
  \hline
 Object & $A$ & $B$ & $C$ & $M(a=1)$ & $M(a=100)$ & $D$\\
 no. & [per cent]  & [per cent] & [per cent] & $[10^{14}\hrmsun]$ & $[10^{14}\hrmsun]$ & [per cent]\\
   \hline
1   &  16.8  &  83.2  &  0.20  &  7.63  &  6.36  &  83.4\\
2   &  18.3  &  81.7  &  0.22  &  6.46  &  5.29  &  81.9\\
3   &  18.7  &  81.3  &  0.50  &  5.76  &  4.71  &  81.8\\
4   &  37.9  &  62.1  &  0.04  &  6.66  &  4.14  &  62.1\\
5   &  25.4  &  74.6  &  0.62  &  5.41  &  4.07  &  75.2\\
6   &  51.6  &  48.4  &  0.21  &  6.64  &  3.23  &  48.6\\
7   &  50.5  &  49.5  &  0.06  &  5.96  &  2.96  &  49.6\\
8   &  26.7  &  73.3  &  0.07  &  3.34  &  2.45  &  73.3\\
9   &  17.2  &  82.8  &  0.67  &  2.88  &  2.40  &  83.4\\
10  &  17.6  &  82.4  &  0.09  &  2.45  &  2.02  &  82.4\\
11  &  29.6  &  70.4  &  0.22  &  2.07  &  1.46  &  70.6\\
  \hline
Mean & 28.2 & 71.8 & 0.26 & & & 72.0\\
Std. dev. & 13.0 & 13.0 & 0.23 & & & 13.1\\
  \hline
  \end{tabular}
  \end{small}
  \label{tabla_D}
\end{table*}

\begin{table*}
  \caption{Quantitative results after application of the
  criterion of \citet{Busha} to 11 objects from the simulation.
  The variables are defined as in Table \ref{tabla_D}.}
  \begin{small}
  \begin{tabular}{| c | c | c | c | c | c | c |}
  \hline
 Object & $A$ & $B$ & $C$ & $M(a=1)$ & $M(a=100)$ & $D$\\
 no. & [per cent]  & [per cent] & [per cent] & $[10^{14}\hrmsun]$ & $[10^{14}\hrmsun]$ & [per cent]\\
   \hline
1   &   7.4  &  92.6  &  9.3   &  6.24  &  6.36  &  101.9\\
2   &   8.7  &  91.3  &  6.0   &  5.44  &  5.29  &   97.2\\
3   &   7.1  &  92.9  & 27.7   &  3.91  &  4.71  &  120.5\\
4   &  10.8  &  89.2  &  2.7   &  4.51  &  4.14  &   91.8\\
5   &   8.4  &  91.6  & 17.2   &  3.74  &  4.07  &  108.8\\
6   &  10.2  &  89.8  & 42.8   &  2.43  &  3.23  &  132.6\\
7   &  15.8  &  84.2  &  5.9   &  3.28  &  2.96  &   90.1\\
8   &  16.3  &  83.7  &  7.2   &  2.69  &  2.45  &   90.8\\
9   &   3.7  &  96.3  & 18.8   &  2.09  &  2.40  &  115.2\\
10  &   5.9  &  94.1  &  4.8   &  2.04  &  2.02  &   98.9\\
11  &  14.9  &  85.1  &  4.3   &  1.64  &  1.46  &   89.4\\
  \hline
Mean & 9.9 & 90.1 & 13.3 & & & 103.4\\
Std. dev. & 4.2 & 4.2 & 12.4 & & &14.3\\
  \hline
  \end{tabular}
  \end{small}
  \label{tabla_B}
\end{table*}

We observe that the purely theoretical criterion is good to
determine the exterior limit of the object since only very few
particles (0.26 per cent of the particles that satisfied the
criterion at $a = 1$) that are outside the criterion today will fall
inside in the late future. In contrast, a significant number of
particles (28 per cent of the particles under the criterion) were
predicted to belong, but finally escaped. Finally, a large number of
particles (72 per cent of the particles under the criterion) were
correctly predicted to belong to the overdensity\footnote{Mean
values from the 11 objects studied.}. Based on this result, we can
assert that the theoretical criterion is adequate to give an upper
bound to the structure extension, but overestimates its mass,
which turns out to be about 72 per cent of the predicted mass.

With the criterion of \citet{Busha}, a greater number of particles
(13 per cent of the particles under the criterion at $a = 1$) fell
outside the criterion today, but ended up inside the object. In
contrast, a smaller number of particles (10 per cent of the
particles that satisfied the criterion) were incorrectly predicted
to belong, while a very large number of particles (90 per cent of
the particles under the criterion) were correctly recognized. Thus,
although it is based on an inconsistent application of the spherical
collapse model, it happens to give a better estimate for the final
object mass, underestimating it by only $\approx 3$ per cent.


It is also important to add that objects 6 and 7 had very complex
cores, being separated into two main concentrations. Of these, we
selected the most massive as the centre of the spherical analysis.
We kept our selection criterion for the centre, under the assumption
that it should be easier to detect the centre of the most massive
concentration when dealing with real observations.

\subsection{Radial velocity predictions}

An important prediction of the spherical collapse model is the
radial velocity of shells falling toward the gravitational
attractor. The velocity information is contained in the velocity
parameter $A$, which depends only on the mass energy density inside
the shell, $\Omega_\mathrm{s}$. This can be calculated numerically
for every shell of the studied objects, yielding the desired radial
velocity.

\begin{figure*}
    \includegraphics[width = 84mm]{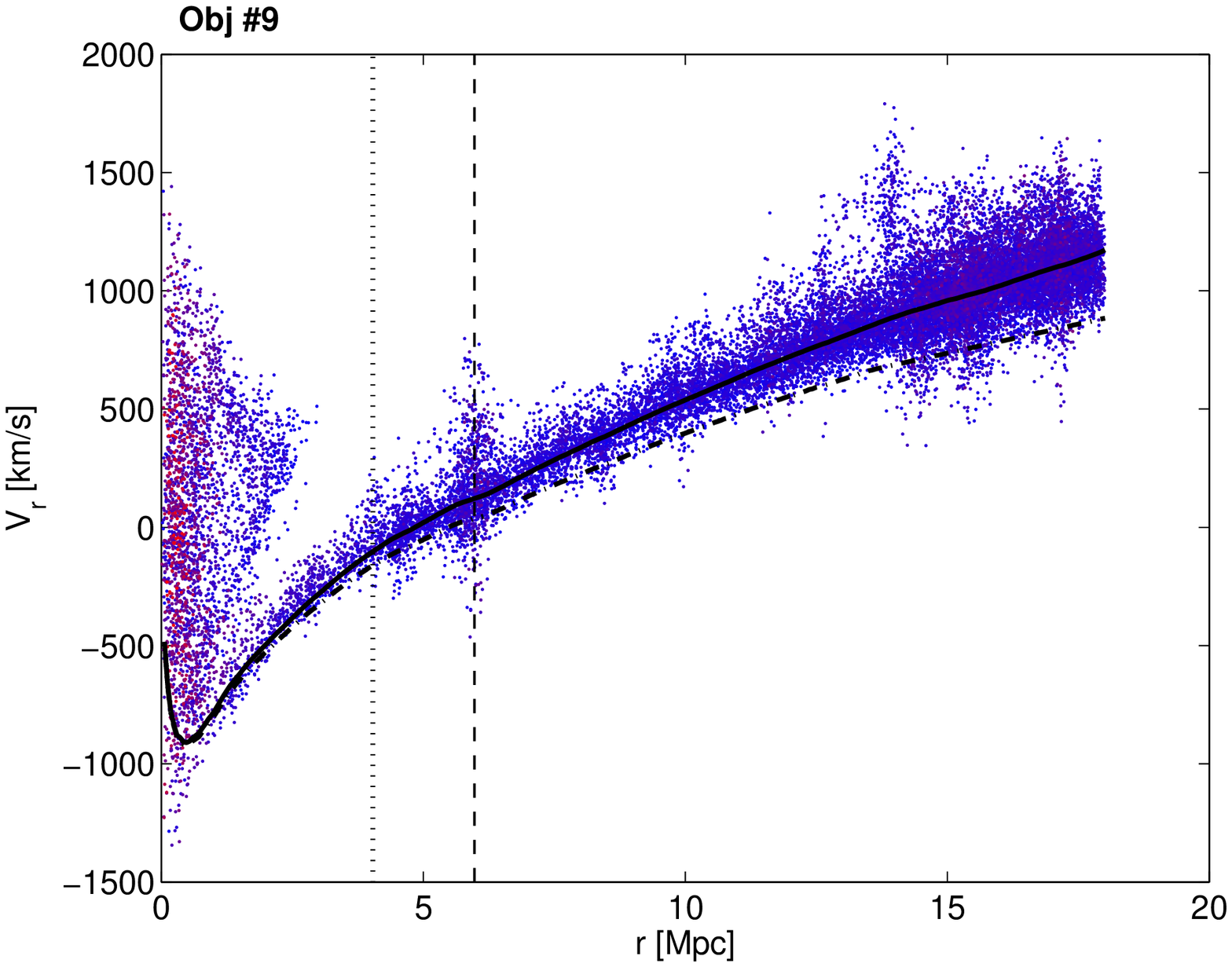}
    \includegraphics[width = 84mm]{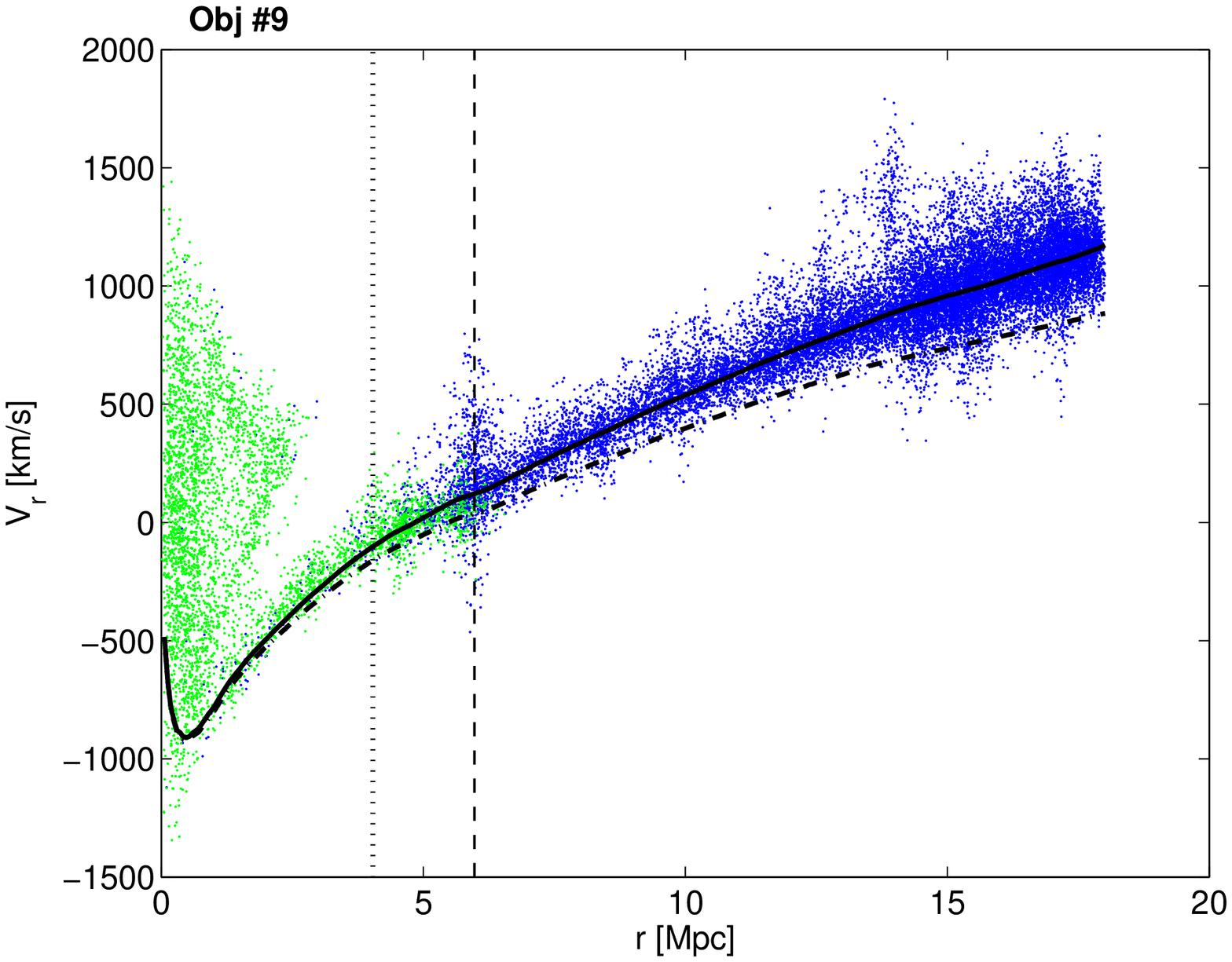}
    \caption{Radial velocities as a function of radius for a mass
    overdensity at $a=1$ for object no. 9. Left-hand panel: colours as a function
    of tangential velocities. Red dots indicate more radial
    velocity and bluer ones less radial velocity. Right-hand panel: colours
    depending on final fate of particles: green particles remain
    bound to the structure, blue particles do not. The vertical
    dashed line shows the radius where
    $\Omega_\mathrm{s} < \Omega_\mathrm{cs} = 2.36$, while the
    vertical dotted line indicates the critical radius according
    to \citet{Busha}. The solid curve is the theoretical
    approximation considering a flat universe with a cosmological
    constant. The dashed curve is the theoretical approximation
    considering a universe without cosmological constant and
    $\Omega_M = 0.3$.}
    \label{fig:Vr_tipo}
\end{figure*}
In Fig. \ref{fig:Vr_tipo}, we plot the radial velocity against
radius, together with the theoretical approximation using the
spherical collapse model with and without cosmological constant (see
\citealt{Maze}, for details about the approximation using the
spherical collapse model without a cosmological constant). For this
analysis, we chose object no.9 because it presented a clear stream of
infalling particles flowing at high speed to the object's core,
cleanly separated from the virialized particles and from particles
currently outflowing after a first pass through the centre, forming
an empty space reaching as close as $2\Mpc$ from the core. Other
objects presented similar overall characteristics, but they did not
show such a clear separation between infalling particles and
virialized ones, probably as a result of earlier virialization or
the presence of substructure. We observe that the spherical model
with cosmological constant is accurate to determine the mean radial
velocity of the infalling part of the cluster particles,
correcting the underestimate of radial velocities at large radius
seen in the spherical model without cosmological constant.

An important observation is that the theoretical velocity profile
follows the infalling particles deep into the core of the structure,
where virialization effects are very important. This result, which
was confirmed on every object we studied, tells us that the
spherical collapse model produces robust predictions of the negative
velocity envelope profile even in highly virialized cores. The
decrease on the predicted velocity very near to the centre is
probably due to the poor resolution of the simulation and to extreme
virialization effects.

\subsection{Perturbations from spherical collapse}
\label{Perturbations From Spherical Collapse} 
\begin{figure}
    \includegraphics[width = 84mm]{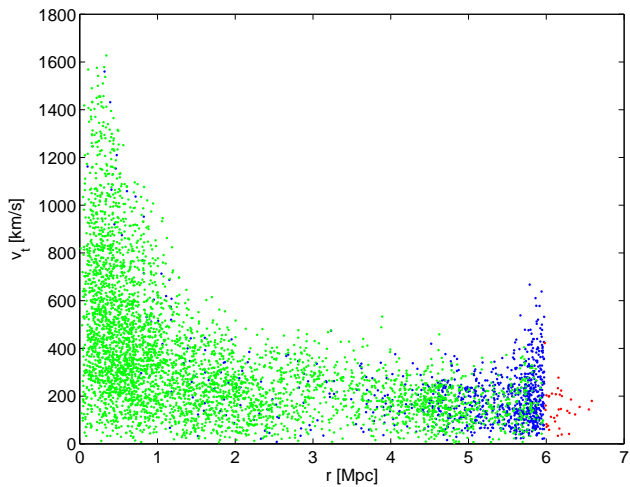}
    \caption{Tangential velocities as a function of radius for object no. 9 at $a=1$.
    Colors depend on the final fate of particles, as given in Fig. \ref{fig:3D_distr}.}
    \label{fig:vt_r}
\end{figure}

In the previous
analyses, we observe that there is a considerable number of
particles that escape from the structure, contradicting the
theoretical prediction. There are two main candidates to be
responsible for this: one is the appearance of tangential velocities
during the contraction, which is also responsible for virialization,
the other is the influence of external structures which can act as
gravitational attractor on external shells. A simple test is to plot
tangential velocities as a function of radius. We would expect that
if there is a clear relation between angular momentum and failure of
the criterion, we would find that particles that contradict the
criterion would have greater tangential velocities than particles
that verify the criterion. This was done for object no. 9 and results
are shown in Fig. \ref{fig:vt_r}.

As seen in Fig. \ref{fig:vt_r}, and as we observed in all other
objects, there is no clear relation between tangential velocity and
probability of escaping. This result is solid enough to say that
angular momentum is not the main reason for the failure of the
criterion, so the hypothesis of external attractors seems more
acceptable. Just as a follow-up observation, we can see in Fig.
\ref{fig:3D_distr} that the particles incorrectly identified as
bound by our criterion (blue) are commonly related to denser regions
outside and close to the critical shell, while the particles
incorrectly identified as unbound (red) are close to mass
concentrations inside the critical shell that finally fall to the
centre. We conclude that the failure of the method is mainly caused
by external overdensities or perturbations from the spherical
distribution of the studied object.

\section{Conclusions}

We have presented a complete discussion of the model of spherical,
gravitational collapse in a flat Universe with a cosmological
constant, applied to estimate the size and mass of bound structures
in the Universe. Within this model, we derive an exact, analytical
equation for the minimum enclosed mass density required for a shell
to remain bound until a very distant future. For the present
cosmological parameters ($\Omega_\mathrm{m}=0.3$,
$\Omega_\Lambda=0.7$), this minimum density is 2.36 times the
critical density of the Universe, as found numerically by
\citet{Tzihong}. This suggests a both physical and practical
criterion for the limits of a bound structure, such as the
previously fairly ill-defined superclusters of galaxies, as the
shell enclosing precisely this density.

The application of the model to simulated data gave encouraging
results, demonstrating first its great ability to find the limits of
structure in the late future, and, second, giving reasonably good
results in determining the limits of bound structures at the present
epoch. On average, 72 per cent of the mass enclosed by the estimated
radius is really bound to the structure, while the mass that,
although bound to the structure, is not enclosed by the radius is
only a 0.3 per cent. For the less rigorous criterion of
\citet{Busha} and other authors, these numbers are 90 and 13 per cent,
giving a substantially better estimate for the object's final mass.
Thus, the sphere defined by our criterion is an outer envelope
enclosing all the particles bound to the structure (and quite a few
more), while that of \citet{Busha} encloses as much mass as will
remain bound to the distant future (leaving about as many bound
particles outside as unbound ones inside).

The spherical collapse model also defines a radial velocity profile
that we will use in our next paper (D\"unner et al., in preparation)
to find the shape of these structures in redshift space, in order to
make them identifiable in redshift surveys. This profile was found
in the simulations to fit the observed velocity profile of infalling
particles well down to deep inside the virialized radius. Thus, even
though the spherical collapse model is intrinsically unstable for
contracting shells, it still gives a reliable performance in a broad
set of radii. Moreover, we observed that the greatest perturbations
from the theoretical model were produced by gravitational
perturbations by external structures or by substructures falling
into the external shells of the structure.

'\section*{Acknowledgments} The authors thank Volker Springel for
generously allowing the use of GADGET2 before its public release.
A.~R. is grateful to Ricardo Demarco, Patricia Ar\'evalo, and Roxana
Contreras for collaboration in unpublished, earlier work on related
subjects that helped stimulate and define the present one. R.~D. and
A.~R. received support from FONDECYT through Regular Project
1020840, and A. M. from the Comit\'e Mixto ESO-Chile. P.~A.~A.
thanks LKBF and the University of Groningen for supporting his visit
to PUC. An anonymous referee and Mat{\'\i}as Carrasco are thanked
for comments that helped improve the manuscript.

\label{lastpage}

\end{document}